\input epsf
%
%
\catcode`\@=11 
\newcount\yearltd\yearltd=\year\advance\yearltd by -1900
%
\def\noblackbox{\overfullrule=0pt}

\def\draftmode{\message{ DRAFTMODE }\def\draftdate{{\rm preliminary draft:
\number\month/\number\day/\number\yearltd\ \ \hourmin}}%
\headline={\hfil\draftdate}\writelabels\baselineskip=20pt plus 2pt minus 2pt
 {\count255=\time\divide\count255 by 60 \xdef\hourmin{\number\count255}
  \multiply\count255 by-60\advance\count255 by\time
  \xdef\hourmin{\hourmin:\ifnum\count255<10 0\fi\the\count255}}}
\def\nolabels{\def\wrlabeL##1{}\def\eqlabeL##1{}\def\reflabeL##1{}}
\def\writelabels{\def\wrlabeL##1{\leavevmode\vadjust{\rlap{\smash%
{\line{{\escapechar=` \hfill\rlap{\sevenrm\hskip.03in\string##1}}}}}}}%
\def\eqlabeL##1{{\escapechar-1\rlap{\sevenrm\hskip.05in\string##1}}}%
\def\reflabeL##1{\noexpand\llap{\noexpand\sevenrm\string\string\string##1}}}
\nolabels
%
\global\newcount\secno \global\secno=0
\global\newcount\meqno \global\meqno=1
\def\newsec#1{\global\advance\secno by1\message{(\the\secno. #1)}
\global\subsecno=0\eqnres@t\noindent{\bf\the\secno. ~#1}
\writetoca{{\secsym} {#1}}\par\nobreak\medskip\nobreak}
\def\eqnres@t{\xdef\secsym{\the\secno.}\global\meqno=1\bigbreak\bigskip}
\def\sequentialequations{\def\eqnres@t{\bigbreak}}\xdef\secsym{}
\global\newcount\subsecno \global\subsecno=0
\def\subsec#1{\global\advance\subsecno by1\message{(\secsym\the\subsecno. #1)}
\ifnum\lastpenalty>9000\else\bigbreak\fi
\noindent{\it\secsym\the\subsecno ~#1}\writetoca{\string\quad
{\secsym\the\subsecno.} {#1}}\par\nobreak\medskip\nobreak}
\def\appendix#1{\global\meqno=1\global\subsecno=0\xdef\secsym{\hbox{#1}}
\bigbreak\bigskip\noindent{\bf #1}
\writetoca{{#1}}\par\nobreak\smallskip\nobreak}
%
%
\def\eqnn#1{\xdef #1{(\secsym\the\meqno)}\writedef{#1\leftbracket#1}%
\global\advance\meqno by1\wrlabeL#1}
\def\eqna#1{\xdef #1##1{\hbox{$(\secsym\the\meqno##1)$}}
\writedef{#1\numbersign1\leftbracket#1{\numbersign1}}%
\global\advance\meqno by1\wrlabeL{#1$\{\}$}}
\def\eqn#1#2{\xdef #1{(\secsym\the\meqno)}\writedef{#1\leftbracket#1}%
\global\advance\meqno by1$$#2\eqno#1\eqlabeL#1$$}
%
%
\global\newcount\refno \global\refno=1
\newwrite\rfile
\def\ref{$^{\the\refno}$\nref}
\def\nref#1{\xdef#1{\the\refno.}\writedef{#1\leftbracket#1}%
\ifnum\refno=1\immediate\openout\rfile=refs.tmp\fi
\global\advance\refno by1\chardef\wfile=\rfile\immediate
\write\rfile{\noexpand\item{#1\ }\reflabeL{#1\hskip.31in}\pctsign}\findarg}
\def\findarg#1#{\begingroup\obeylines\newlinechar=`\^^M\pass@rg}
{\obeylines\gdef\pass@rg#1{\writ@line\relax #1^^M\hbox{}^^M}%
\gdef\writ@line#1^^M{\expandafter\toks0\expandafter{\striprel@x #1}%
\edef\next{\the\toks0}\ifx\next\em@rk\let\next=\endgroup\else\ifx\next\empty%
\else\immediate\write\wfile{\the\toks0}\fi\let\next=\writ@line\fi\next\relax}}
\def\striprel@x#1{} \def\em@rk{\hbox{}}
\def\lref{\begingroup\obeylines\lr@f}
\def\lr@f#1#2{\gdef#1{\ref#1{#2}}\endgroup\unskip}

\def\addref#1{\immediate\write\rfile{\noexpand\item{}#1}} 
\def\startrefs#1{\immediate\openout\rfile=refs.tmp\refno=#1}
\def\xref{\expandafter\xr@f}\def\xr@f#1.{#1}
\def\cite{\expandafter\cxr@f}\def\cxr@f#1.{$^{#1}$}
\def\xcite{\expandafter\xcxr@f}\def\xcxr@f#1.{{#1}}
\def\refs#1{\count255=1$^{\r@fs #1{\hbox{}}}$}
\def\r@fs#1{\ifx\und@fined#1\message{reflabel \string#1 is undefined.}%
\nref#1{need to supply reference \string#1.}\fi%
\vphantom{\hphantom{#1}}\edef\next{#1}\ifx\next\em@rk\def\next{}%
\else\ifx\next#1\ifodd\count255\relax\xref#1\count255=0\fi%
\else#1\count255=1\fi\let\next=\r@fs\fi\next}
\newwrite\lfile
{\escapechar-1\xdef\pctsign{\string\%}\xdef\leftbracket{\string\{}
\xdef\rightbracket{\string\}}\xdef\numbersign{\string\#}}

\def\writestop{\def\writestoppt{\immediate\write\lfile{\string\pageno%
\the\pageno\string\startrefs\leftbracket\the\refno\rightbracket%
\string\def\string\secsym\leftbracket\secsym\rightbracket%
\string\secno\the\secno\string\meqno\the\meqno}\immediate\closeout\lfile}}
\def\writestoppt{}\def\writedef#1{}
\def\seclab#1{\xdef #1{\the\secno}\writedef{#1\leftbracket#1}\wrlabeL{#1=#1}}
\def\subseclab#1{\xdef #1{\secsym\the\subsecno}%
\writedef{#1\leftbracket#1}\wrlabeL{#1=#1}}
\newwrite\tfile \def\writetoca#1{}
\def\leaderfill{\leaders\hbox to 1em{\hss.\hss}\hfill}
\def\writetoc{\immediate\openout\tfile=toc.tmp
   \def\writetoca##1{{\edef\next{\write\tfile{\noindent ##1
   \string\leaderfill {\noexpand\number\pageno} \par}}\next}}}

%
\catcode`\@=12 
%
%
\font\abssl=cmsl10 scaled 833
\font\absrm=cmr10 scaled 833 \font\absrms=cmr7 scaled  833
\font\absrmss=cmr5 scaled  833 \font\absi=cmmi10 scaled  833
\font\absis=cmmi7 scaled  833 \font\absiss=cmmi5 scaled  833
\font\abssy=cmsy10 scaled  833 \font\abssys=cmsy7 scaled  833
\font\abssyss=cmsy5 scaled  833 \font\absbf=cmbx10 scaled 833
\skewchar\absi='177 \skewchar\absis='177 \skewchar\absiss='177
\skewchar\abssy='60 \skewchar\abssys='60 \skewchar\abssyss='60
\def\abstractfont{\def\rm{\fam0\absrm}
\textfont0=\absrm \scriptfont0=\absrms \scriptscriptfont0=\absrmss
\textfont1=\absi \scriptfont1=\absis \scriptscriptfont1=\absiss
\textfont2=\abssy \scriptfont2=\abssys \scriptscriptfont2=\abssyss
\textfont\itfam=\absi \def\it{\fam\itfam\absi}
\textfont\slfam=\abssl \def\sl{\fam\slfam\abssl}
\textfont\bffam=\absbf \def\bf{\fam\bffam\absbf}\rm}
\font\ftsl=cmsl10 scaled 833
\font\ftrm=cmr10 scaled 833 \font\ftrms=cmr7 scaled  833
\font\ftrmss=cmr5 scaled  833 \font\fti=cmmi10 scaled  833
\font\ftis=cmmi7 scaled  833 \font\ftiss=cmmi5 scaled  833
\font\ftsy=cmsy10 scaled  833 \font\ftsys=cmsy7 scaled  833
\font\ftsyss=cmsy5 scaled  833 \font\ftbf=cmbx10 scaled 833
\skewchar\fti='177 \skewchar\ftis='177 \skewchar\ftiss='177
\skewchar\ftsy='60 \skewchar\ftsys='60 \skewchar\ftsyss='60
\def\footnotefont{\def\rm{\fam0\ftrm}
\textfont0=\ftrm \scriptfont0=\ftrms \scriptscriptfont0=\ftrmss
\textfont1=\fti \scriptfont1=\ftis \scriptscriptfont1=\ftiss
\textfont2=\ftsy \scriptfont2=\ftsys \scriptscriptfont2=\ftsyss
\textfont\itfam=\fti \def\it{\fam\itfam\fti}%
\textfont\slfam=\ftsl \def\sl{\fam\slfam\ftsl}%
\textfont\bffam=\ftbf \def\bf{\fam\bffam\ftbf}\rm}
\font\ninerm=cmr9 \font\sixrm=cmr6 \font\ninei=cmmi9 \font\sixi=cmmi6
\font\ninesy=cmsy9 \font\sixsy=cmsy6 \font\ninebf=cmbx9
\font\nineit=cmti9 \font\ninesl=cmsl9 \skewchar\ninei='177
\skewchar\sixi='177 \skewchar\ninesy='60 \skewchar\sixsy='60
\def\ninepoint{\def\rm{\fam0\ninerm}
\textfont0=\ninerm \scriptfont0=\sixrm \scriptscriptfont0=\fiverm
\textfont1=\ninei \scriptfont1=\sixi \scriptscriptfont1=\fivei
\textfont2=\ninesy \scriptfont2=\sixsy \scriptscriptfont2=\fivesy
\textfont\itfam=\ninei \def\it{\fam\itfam\nineit}\def\sl{\fam\slfam\ninesl}%
\textfont\bffam=\ninebf \def\bf{\fam\bffam\ninebf}\rm}
%
%

\vsize=7.0truein
\hsize=4.7truein
\baselineskip 12truept plus 0.5truept minus 0.5truept
\hoffset=0.5truein
\voffset=0.5truein
\def\1{\;1\!\!\!\! 1\;}

\def\neath#1#2{\mathrel{\mathop{#1}\limits_{#2}}}

\def\epm#1#2{\hbox{${+#1}\atop {-#2}$}}
\def\neath#1#2{\mathrel{\mathop{#1}\limits_{#2}}}
\def\gsim{\mathrel{\rlap{\lower4pt\hbox{\hskip1pt$\sim$}}
    \raise1pt\hbox{$>$}}}         

\def\frac#1#2{{{#1}\over {#2}}}

\def\smallfrac#1#2{\hbox{${{#1}\over {#2}}$}}

\def\as{\alpha_s}

\catcode`@=11 
\def\slash#1{\mathord{\mathpalette\c@ncel#1}}
 \def\c@ncel#1#2{\ooalign{$\hfil#1\mkern1mu/\hfil$\crcr$#1#2$}}
\def\lsim{\mathrel{\mathpalette\@versim<}}
\def\gsim{\mathrel{\mathpalette\@versim>}}
 \def\@versim#1#2{\lower0.2ex\vbox{\baselineskip\z@skip\lineskip\z@skip
       \lineskiplimit\z@\ialign{$\m@th#1\hfil##$\crcr#2\crcr\sim\crcr}}}
\catcode`@=12 

\def\PR{{\it Phys.~Rev.~}}
\def\PRL{{\it Phys.~Rev.~Lett.~}}
\def\NP{{\it Nucl.~Phys.~}}

\def\PL{{\it Phys.~Lett.~}}

\def\ZP{{\it Zeit.~Phys.~}}

\def\vol#1{{\bf #1}}\def\vyp#1#2#3{\vol{#1} (#2) #3}

\noblackbox
\tolerance=10000
\hfuzz=5pt
\pageno=0\nopagenumbers\tolerance=10000\hfuzz=5pt
\line{\hfill {\tt hep-ph/9610238}}
\line{\hfill DFTT 57/96}
\vskip 12pt
\centerline{\bf POLARIZED STRUCTURE FUNCTIONS:}
\centerline{\bf A STATUS REPORT}
\vskip 24pt
\centerline{Stefano Forte}
\vskip 10pt
\centerline{\it INFN, Sezione di Torino}
\centerline{\it via P. Giuria 1, I-10125 Torino, Italy}
\vskip 36pt
{\narrower\baselineskip 10pt
\centerline{\bf Abstract}
\medskip\noindent
We review the present status of
polarized structure
functions measured in deep-inelastic scattering. We discuss
the $x$ and $Q^2$ dependence of the
structure function $g_1$, and how it can be used to
 test perturbative
QCD at next-to-leading order and beyond. We summarize the current
knowledge of polarized parton distributions,
in particular the determination of the first
moment of the quark and gluon distributions and the axial charge of the
nucleon. We critically examine
what future experiments could teach us on the polarized structure
of the nucleon.
\smallskip}
\vskip 36 pt
\centerline{Invited plenary talks given at}
\medskip
\centerline{ {\it PANIC 96}}
\centerline{ Williamsburg, Virginia, May 1996}
\smallskip
\centerline{and}
\smallskip
\centerline{ {\it SPIN 96}}
\centerline{Amsterdam, the Netherlands, September 1996}
\bigskip
\centerline{\it to be published in the proceedings}
\vfill
\line{October 1996\hfill}
\eject
\footline={\hss\tenrm\folio\hss}
\centerline{\bf POLARIZED STRUCTURE FUNCTIONS:}
\centerline{\bf A STATUS REPORT}
\bigskip\bigskip
\centerline{STEFANO FORTE}
\smallskip
\centerline{\it INFN, Sezione di Torino}
\centerline{\it via P. Giuria 1, I-10125 Torino, Italy}
\smallskip

\bigskip
{\abstractfont\baselineskip 10 pt
\advance\leftskip by 36truept\advance\rightskip by 36truept\noindent
We review the present status of
polarized structure
functions measured in deep-inelastic scattering. We discuss
the $x$ and $Q^2$ dependence of the
structure function $g_1$, and how it can be used to
 test perturbative
QCD at next-to-leading order and beyond. We summarize the current
knowledge of polarized parton distributions,
in particular the determination of the first
moment of the quark and gluon distributions and the axial charge of the
nucleon. We critically examine
what future experiments could teach us on the polarized structure
of the nucleon.
\smallskip}

\baselineskip 12truept plus 0.5truept minus 0.5truept
\medskip
\goodbreak
\newsec{\bf From structure functions to polarized partons.}
\nobreak
Structure functions measured in deep-inelastic scattering
provide the cleanest access to the structure of the nucleon, as
probed in hard processes, and understood in perturbative QCD.
In the polarized case only very recently have theory and experiment
progressed to the point that it is possible to extract
from experiment physically
meaningful information. Indeed, it is now possible
to measure the  nucleon matrix elements of quark
and gluon operators, and compare experimental results
on parton distributions  with the
$x$ and $Q^2$ dependence predicted by
perturbative QCD at next-to-leading order (NLO).

\nref\blois{S.~Forte, in  ``Frontiers in Strong Interactions'', proc.
 of the 7th Rencontres de Blois, M.~Haguenauer, ed. (Editions Fronti\`eres,
Paris, 1996), {\tt hep-ph/9511345}.}
\nref\gatlin{S.~Forte, in  ``Radiative Corrections: Status and Outlook'',
B.~F.~L.~Ward, ed. (World Scientific, Singapore 1995),
 {\tt hep-ph/9409416}.}
\nref\guidoer{G.~Altarelli, in ``The challenging questions'',
Proc. of the 1989 Erice School,
A.~Zichichi, ed. (Plenum, New York, 1990).}
Here we will chart the current status of this knowledge, by
summarizing how recent data test perturbative
QCD, reviewing what we have learnt on the polarized parton structure
of the nucleon, and discussing what future experiments could teach us.
In Sect.~2 we will review recent developments related to the study of the
Bjorken sum rule, which singles
out the hard structure of perturbative QCD. In Sect.~3 we will
discuss the behavior of the structure
function $g_1(x,Q^2)$ in next-to-leading order (NLO) in the $(x,Q^2)$ plane,
with special regard to the small-$x$ region.
In Sect.~4  we will review the current knowledge
of polarized parton distributions, and in
particular their first moments, which are related to the
parton interpretation of the nucleon spin. In Sect.~5 we will finally
compare the information which could be in principle obtained on
polarized partons with current and future experimental
prospects.\footnote*{\footnotefont\baselineskip 10 truept
For a more detailed introduction to the aspects of the theory of the polarized
structure function $g_1(x,Q^2)$ relevant to this paper see
e.g. ref.~\xref\blois, whose notation and conventions we will follow;
a
general discussion of the phenomenology of $g_1$ is e.g. in ref.~\xref\gatlin.
For a comprehensive introduction to the problems related to the determination
of the first moment of $g_1$ (the ``spin crisis'') see ref.~\xref\guidoer.
\vskip-.2truecm}

\goodbreak
\medskip
\newsec{\bf The Bjorken sum rule.}
\nobreak
The nonsinglet first moment of the structure function $g_1$ has the peculiar
feature
of being related to the matrix element of a conserved current
through a nontrivial coefficient function:
\eqn\nonsing
{\Gamma^{\rm NS}_1(Q^2)\equiv\int_0^1\!dx\,g^{\rm NS}_1(x,Q^2)={1\over 2}
C_{\rm NS} [\alpha_s(Q^2)] a_{\rm NS} .}
 This implies that its
measurement probes the hard part of the total
polarized inclusive nonsinglet
$\gamma^*$-$p$ cross section, which is proportional to the coefficient function
$C_{\rm NS} (Q^2)$,
up to a scale-independent coefficient, the nucleon matrix element of
the nonsinglet axial current
$a_{\rm NS}$,\footnote\dag{\footnotefont\baselineskip 10 truept
Strictly speaking, $a_{\rm NS}$ is only scale independent far from
heavy quark thresholds, otherwise it acquires a scale dependence
due to the appearance of a new heavy quark contribution as the
threshold is passed.\vskip1pt} which incorporates all the
soft, target-dependent
physics. Hence, determining $C_{\rm NS} (Q^2)$ at various scales probes the
perturbative computation of this cross-section, which is under theoretical
control and indeed has been accomplished\ref\larns{
S.~G.~Gorishny and S.~A.~Larin, {\it
Phys. Lett.} {\bf B172} (1986) 109;
S.~A.~Larin
and J.~A.~M.~Vermaseren, {\it Phys. Lett.} {\bf B259} (1991) 345.}
up to order $\alpha_s^3$.

In the particular case of the isotriplet
combination the current matrix element can in turn be related, using isospin
algebra, to that which governs nucleon $\beta$ decay, and thus expressed in
terms of the corresponding decay constant $g_A$:
\eqn\tripchar{a^{I=1}= {1\over 6} \langle p,s|{j^\mu_5}^{I=1}|p,s\rangle
{s_\mu\over M}={1\over 6} g_A}
where $M$ and $s^\mu$ are the nucleon mass and spin and the
axial current is $j^\mu_i=\bar\psi\gamma^\mu\gamma_5\lambda_i\psi$, $\lambda_i$
being an isospin matrix. This gives an absolute prediction
for the isotriplet first moment (Bjorken sum rule): testing the
normalization of this prediction tests isospin symmetry in this channel,
and measuring its scale dependence tests perturbative QCD, in particular
the size and running of the strong coupling $\alpha_s(Q^2)$.

\topinsert
\epsfysize=10truecm\vskip-2.5truecm\hskip-.5truecm
\hfil\epsfbox{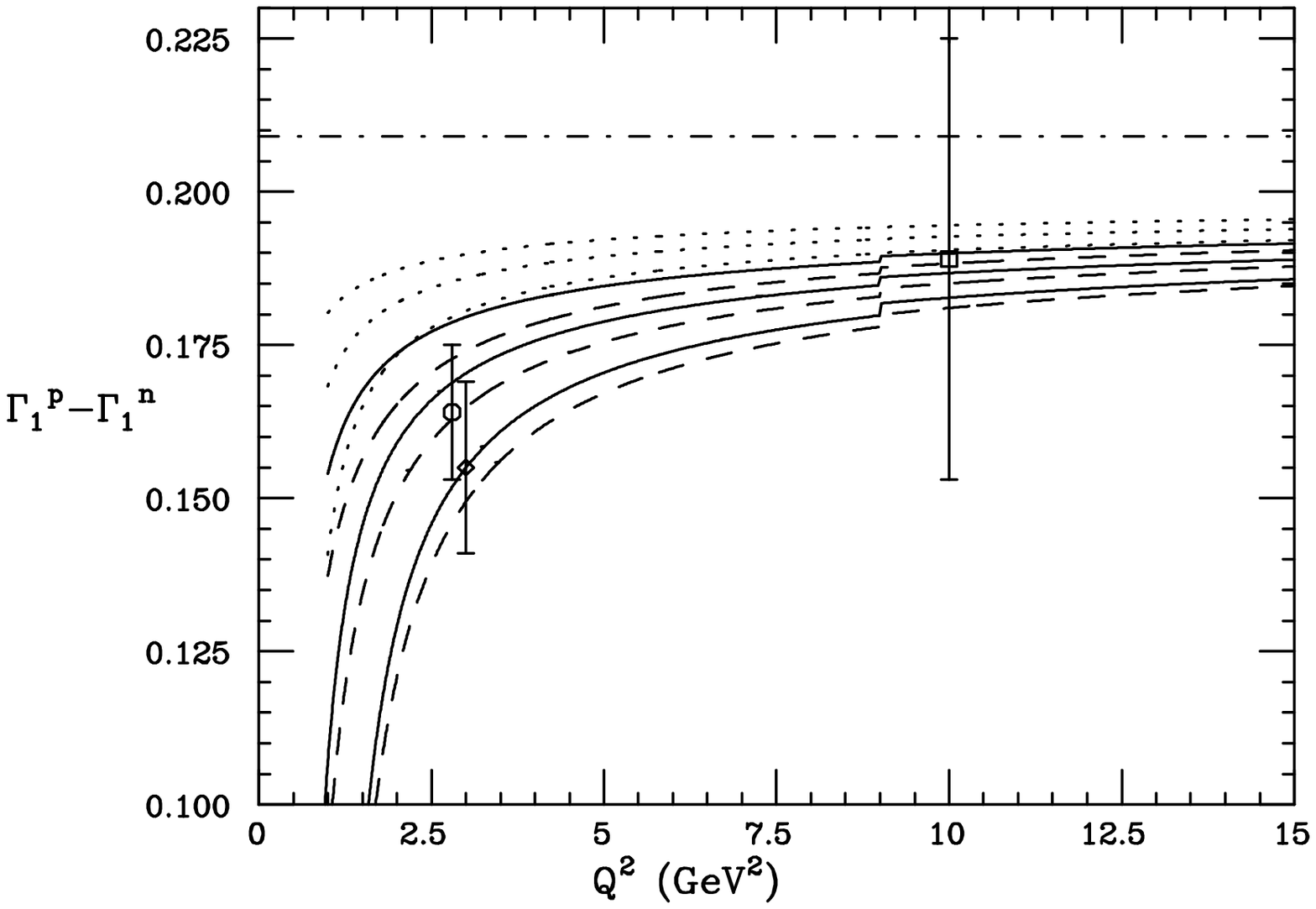}\hfil\vskip-3.truecm
\bigskip\noindent{\abstractfont\baselineskip=10truept
\noindent \nref\smcrm{A.~Witzmann, for the SMC Collab.,
talk at ``DIS 96'', Rome, April 1996.}\nref\slacrm{S.~Rock, for the
E142 and E143 Collab., talk at
``DIS 96'', Rome, April 1996.}\nref\elkar{J.~Ellis and M.~Karliner,
{\it Phys. Lett.}
{\bf B341} (1995) 397; {\tt hep-ph/9601280}.}
\hskip-12pt Fig.~1:
Scale dependence of the Bjorken sum rule up to order $\alpha_s$
(dotted), $\alpha_s^3$ (solid) and with higher twist corrections (dashed).
The three sets of curves correspond to
$\alpha_s(M_Z)=0.119\pm0.006$.
The data points are from ref.~\xref\smcrm\ (square), \xref\slacrm\
(diamond) and  the global average of
ref.~\xref\elkar\ (circle) (slightly offset from $Q^2$=3~GeV$^2$
to improve readability).\vskip1pt}\vskip-.4truecm
\smallskip
\medskip
\endinsert
The predicted scale dependence of the sum rule
(with\ref\guias{G.~Altarelli, in ``DIS 95'',
J.~F.~Laporte and Y.~Sirois, eds. (Ecole Polytechnique, Paris 1996).}
$\alpha_s(M_Z)=0.119\pm0.006$)
is shown in Fig.~1, which demonstrates
its sensistivity both to the value of the strong coupling, and to higher order
corrections. Notice the jump at the charm threshold, which
is due to the explicit $n_f$
dependence of the coefficient function at order $\alpha_s^2$ and beyond
(the discontinuity
would be smoothened in a more refined treatment of thresholds).
Recently, both the the SMC\cite\smcrm\ and
the E142-E143 collaboration\cite\slacrm\ have released
preliminary determinations of the isotriplet first moment,
repectively obtained at $10$~GeV$^2$
combining proton and deuteron data,
and at $3$~GeV$^2$  also  including $^3$He data;
the results are manifestly
consistent with the QCD
prediction. Apparent from Fig.~1 is also the much greater sensitivity
of the sum rule to higher order corrections and the value
of $\alpha_s$ when the scale is low. It follows that a measurement
of equal accuracy leads a priori to a more stringent test of
the sum rule (or, equivalently, to a better determination of $\alpha_s$)
if performed at a low scale, but it is then affected by a larger theoretical
uncertainty related to lack of knowledge of higher order corrections.

In fact, as we will discuss in Sect.~3 and 4, the experimental
values of the first moment are themselves obtained by evolving to a common
scale data taken at several values of $Q^2$, and then extrapolating
to $x=0$ and $x=1$. The evolution  is done\refs{\smcrm,\slacrm}
by means of an approximate procedure (which at best approximates NLO
evolution), involving a systematic error
which is neglected, but is likely to be large at  low scales:
it follows that the apparent
greater precision of the low $Q^2$ data point in fig.~1 is
in part due to the neglect of this systematics.\ref\bfr{R.~D.~Ball,
S.~Forte and G.~Ridolfi, {\it Nucl}. {\it Phys}.
{\bf B 444} (1995) 287.}
If, disregarding these   complications,
the published values of the first moment
are combined (at a common scale), one may
arrive at a determination of the sum rule which taken at face value
is very precise, and would determine $\alpha_s$ to competitive
accuracy,\cite\elkar\
 (see  Fig.~1). This can only be correct if
the aforementioned systematic error cancels
out in the isotriplet combination; which
is presumably true to a certain extent, but has not been quantitatively
established. An example of unavoidable
uncertainty is that
related to the treatment of the charm threshold, which, as shown
in fig.~1, is at least as large as the $O(\alpha^3)$ correction
 to $C_{\rm NS}(Q^2)$.

The perturbative computation of $C_{\rm NS}(Q^2)$ is in fact extended
to high enough order that one may wonder whether the asymptotic
nature of perturbation theory in QCD may already manifest itself.
Specifically, the asymptotic series for $C_{\rm NS}(Q^2)$
can be summed by the Borel method but
only up to an ambiguity of order $1\over Q^2$: one can then try to
estimate the size of this ambiguity, either by approximate
computation of the high-order behavior of the
series,\ref\stein{M.~Meyer-Herrmann et al., {\tt hep-ph/9605229}.}
or on the basis of the terms which are already known (extrapolated
with the method of Pad\'e approximants).\ref\elkarpad{J.~Ellis et al.,
{\it Phys. lett.} {\bf B366} (1996) 268.} This ambiguity must
cancel against an equal and opposite ambiguity in the contribution
$c_{\rm HT}$
to the sum rule from higher twist operators (such as $\bar \psi \gamma^\mu
\gamma_5\psi F_{\mu\nu}$):  $\Gamma^{I=1}={1\over 6}
\left[ C_{\rm NS} (Q^2) g_A+
{C_{\rm HT}\over Q^2}+O(1/Q^4)\right]$.
The estimates of ref.s~\xref\stein,\xref\elkarpad\
lead to a value of the ambiguity which is of the same size as
the contributions of such operators computed with QCD sum rule methods,
which give
$C_{\rm HT}\approx-0.01$~GeV$^2$.\ref\lech{L.~Mankiewicz, E.~Stein
and A.~Sch\"afer,
{\tt hep-ph/9510418}.} This suggests that even though the sum rule
estimates are of the right order of magnitude, their
actual size is entirely uncertain.
The logarithmic corrections to the scale dependence of the twist four, spin
one operators which contribute to the first moment of $g_1$
have also been computed recently.\ref\htscal{H.~Kawamura et al.,
{\tt hep-ph/9603338}.}
\goodbreak
\medskip
\newsec{\bf Parton distributions at next-to-leading order}
\nobreak
The moments of $g_1(N,Q^2)=\int_0^1\!dx\, x^{N-1} g_1(x,Q^2)$
generically depend on scale both because of the
$\alpha_s$ dependence of coefficient functions, and because of the
scale dependence of the matrix elements of quark and gluon operators
to which they are related by
\eqn\gone
{\eqalign{g_1(N,Q^2)=&
\smallfrac{\langle e^2\rangle}{2}[
C_{\rm NS}(N,\alpha_s)\Delta q_{\rm NS}(N,Q^2)
+\cr
&\> C_{\rm S}(N,\alpha_s) \Delta\Sigma(N,Q^2)
+ 2n_f C_g(N,\alpha_s)\Delta g(N,Q^2)],\cr}}
where
$\Delta q_{\rm NS}=\sum_{i=1}^{n_f}
\left(\smallfrac{e_i^2}{\langle e^2\rangle}-1\right)
(\Delta q_i+\Delta\bar q_i)$ and
$\Delta \Sigma =\sum_{i=1}^{n_f} (\Delta q_i+\Delta\bar q_i)$
(with $\langle e^2\rangle=\smallfrac{1}{n_f}\sum_{i=1}^n e^2_i$)
are respectively the nonsinglet and singlet combinations
of moments of the quark distributions.
The scale dependence of the polarized quark and gluon distributions
is in turn governed by the evolution equations
\eqn\aps{\eqalign{\frac{d}{dt}\Delta q_{\rm NS}
&=\frac{\as(t)}{2\pi} \gamma_{\rm NS}^N\Delta q_{\rm NS}\cr
\frac{d}{dt}\pmatrix{\Delta\Sigma\cr \Delta g\cr}&=
\frac{\as(t)}{2\pi}
\pmatrix{\gamma^{N}_{qq}& 2 n_f
\gamma^{N}_{qg}\cr \gamma^{N}_{gq}&
\gamma^{N}_{gg}\cr}
\pmatrix{\Delta\Sigma\cr \Delta g\cr}.\cr}}

The physically relevant moments of quark
and gluon distributions can thus be extracted from the measurement
of the moments of $g_1$ at a pair of different scales.
However, the moments of $g_1$ cannot be determined directly
because the structure function can only be measured in a finite
range of $x$ ($x\to0$ corresponds to the energy of the $\gamma^*$-nucleon
collision going to infinity); furthermore, present-day experiments only
determine $g_1$ along a curve $Q(x)$ (rather than at a fixed scale)
with $Q$ increasing with $x$.
This implies that the moments of $g_1$ can only be extracted indirectly,
through an analysis of its $x$ and $Q^2$
dependence: all the available experimental information is then summarized
in a set of polarized parton distributions at a reference scale, from
which $g_1$ at all $x$ and $Q^2$ are found by solving the evolution
equations and taking (and inverting) moments.
The potential accuracy of such an analysis has
substantially improved recently since the complete
set of anomalous dimensions $\gamma_{ij}^N$ has now
been  determined up to
NLO.\ref\nlo{R.~Mertig and W.~L.~van~Neerven, \ZP\vyp{C70}{1996}{637};
W.~Vogelsang, \PR\vyp{D54}{1996}{2023}.}

Given the need to extrapolate the data to $x=0$ in order to compute
moments,
it is particularly interesting to study the implications of perturbative
evolution for the small $x$ behavior of $g_1$, which follows from the
behavior of the anomalous dimensions $\gamma_{ij}^N$ eq.~\aps\ around their
rightmost singularity in $N$ space. This is located at $N=0$ and
is on general grounds of the form\ref\gross{D.~Gross, in ``Methods
in field theory'', R.~Balian and J.~Zinn-Justin, eds. (North-Holland,
Amsterdam, 1976).}
$\alpha_s\gamma\neath
\sim{N\to0} ({\alpha_s/N})^{2n+1}$ at $n$-th perturbative order.
Already at LO this implies that any starting parton distributions leads
to a growth of $|g_1(x,Q^2)|$ at small $x$ and
large $Q^2$;\ref\rise{M.~A.~Ahmed and
G.~G.~Ross, {\it Phys. Lett.} {\bf B56} (1975)
385; M.~B.~Einhorn and J.~Soffer, {\it Nucl. Phys.} {\bf B74} (1986) 714;
A.~Berera, {\it Phys. Lett.} {\bf B293} (1992)
445.} in fact in this limit the ratio of $\Delta  g$ and $\Delta \Sigma$
is fixed and negative, so that $g_1<0$ asymptotically at small $x$ and
large $Q^2$
for any reasonable input parton
distribution.\cite\bfr\
The generic form of the rise at $n$-th perturbative
order is then
\eqn\sxasy{\Delta f\sim {1\over (\xi\zeta)^{1/4}}
e^{2 \gamma_f\sqrt{\xi\zeta}}\left(1+\sum_{i=1}^n\epsilon_f^i
\left(\sqrt{\xi\over\zeta}\right)^{2i+1} \alpha_s^i\right),}
where $\xi\equiv\ln\smallfrac{x_0}{x}$,
$\zeta\equiv\ln\smallfrac{\alpha_s(Q_0^2)}{\alpha_s(Q^2)}$,
and $\Delta f$ denotes either the nonsinglet quark distribution,
or  the two linear combinations of the gluon and singlet quark
distributions which correspond to eigenvectors of the small $N$
anomalous dimension matrix. Notice that this rise contradicts the
Regge expectation\ref\heim{R.~L.~Heimann, {\it Nucl. Phys.} {\bf B64} (1973)
429.} that $g_1$ should at most be constant at small $x$.

The coefficients $\epsilon_f$ in eq.~\sxasy\ at NLO can be
extracted\ref\bfra{R.D.~Ball,
S.~Forte and G.~Ridolfi, \PL\vyp{B378} {1996} {255}.} from the full
two-loop anomalous dimension;\cite\nlo\ beyond NLO they
have been obtained at  fixed coupling both in the
nonsinglet\nref\kl{R.~Kirschner and L.~Lipatov,
\NP\vyp{B213}{1983}{122}.}\nref\ermns{J.~Bartels,
B.~I.~Ermolaev and M.~G.~Ryskin,
{\it Z. Phys.} {\bf C70} (1996) 273.}\refs{\kl,\ermns} and
singlet\ref\erms{J.~Bartels, B.~I.~Ermolaev and M.~G.~Ryskin,
{\tt hep-ph/9603204}.} case
on the basis of the analysis of the leading double logarithmic singularities.
These results (unlike their unpolarized counterparts) are not supported by
suitable factorization theorems,  even though
they do agree with the NLO result\cite\bfra.
If correct, they lead to the prediction that the
series in eq.~\sxasy\ exponentiates, thus leading to a power-like
rise of $g_1\sim x^{-\lambda}$.
The value of $\lambda$ cannot be computed precisely
since it depends on $\alpha_s$, which
is assumed to be constant in these computations; taking $\alpha_s\sim 0.2$
(which corresponds to a scale of roughly 10~GeV$^2$) would lead
to $\lambda\approx 0.5$ in the nonsinglet and $\lambda\gsim 1$
in the singlet case, implying that the first moment of $g_1$
would then diverge. This casts some doubts on the applicabilty
of the analysis, at least in its naive form. However, it suggests
that a strong rise of $g_1$, which might be of considerable phenomenological
relevance, should be generated by perturbative QCD. The all-order
analysis confirms the LO result that
this rise will preserve the sign of the
nonsinglet, but lead to a negative singlet contribution
(which asymptotically
dominates $g_1$ at small $x$)
even if $g_1$ is positive at the starting scale.
\goodbreak
\medskip
\newsec{\bf Phenomenology of $g_1$.}
\nobreak

\nref\smcp{SMC Collab.,
\PL\vyp{B329}{1994}{399}.} \nref\slacp{E143
Collab.,  \PRL\vyp{74}{1995}{346}.}
\nref\smcd{SMC Collab.,
\PL\vyp{B357}{1995}{248}.}
\nref\slacd{E143
Collab.,
\PRL\vyp{75}{1995}{25}.}\nref\slacsv{E143
Collab., \PL\vyp{B364}{1995}{61}.}
\topinsert
\vskip-4.truecm\epsfysize=13truecm
\hfil\epsfbox{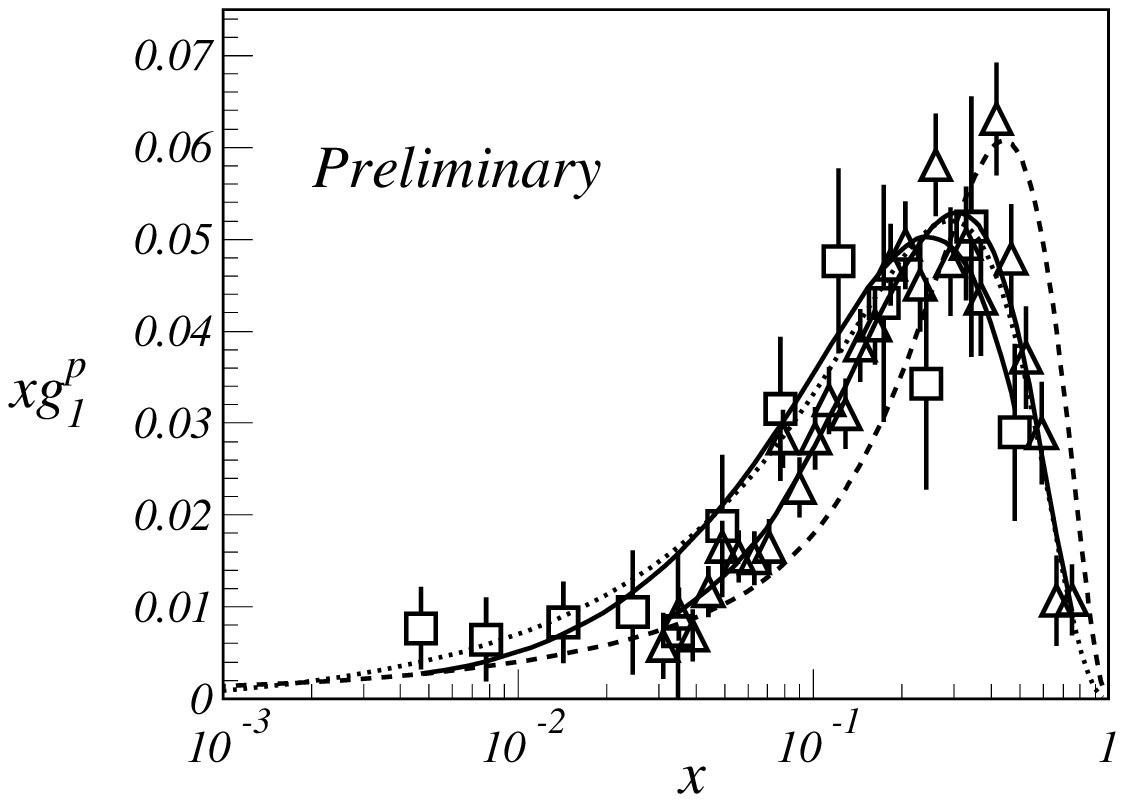}\hfil\vskip-5.truecm
\bigskip\noindent{\abstractfont\baselineskip=10truept
\hskip-8pt Fig. 2: Plot of $xg_1^p$. The squares are SMC
data (preliminary)\refs{\smcp,\smcrm} and the
diamonds E143 data.\cite\slacp\
The curves are the results of a NLO fit\ref\newfit{R.~D.~Ball, A.~Deshpande,
S.~Forte, V.~Hughes, J.~Lichtenstadt and G.~Ridolfi, {\it unpublished}.} to
all\refs{\smcp-\slacsv,\smcrm}
available data: long dashes: $Q^2$=1~GeV$^2$,
short dashes: $Q^2$=10~GeV$^2$, solid: $Q^2=Q^2(x)$ corresponding to the
two data sets.
\vskip1pt}\vskip-.4truecm
\smallskip
\medskip
\endinsert
The scaling violations displayed by current data for $g_1^p$ are shown
in Fig.~2: what might appear as a systematic discrepancy of the
two data sets\refs{\smcp,\slacp}
is in fact a consequence of the fact that in each $x$ bin
the SMC data are taken at a higher value of
$Q^2$ than the E143 data.
These scaling violations are partly due to scale dependence
of the asymmetry $A_1=g_1/F_1$, i.e. they would be
significantly smaller if the
asymmetry were scale independent.
Thus, even though the individual experiments, because
of their limited kinematic coverage,
cannot measure\refs{\slacsv,\smcrm}  the scale
dependence of the asymmetry directly, the data do show evidence for this
scale dependence when  the two data sets are combined.

\smallskip
\subsec{\it Parton distributions}
The observation of sizable scaling violations, together with the availability
of proton\refs{\smcp,\slacp}
and deuteron\refs{\smcd,\slacd} data which allow  the
disentangling of singlet and nonsinglet contributions, make a
full NLO determination
of polarized parton distributions
possible.\nref\gr{M.~Gl\"uck et al., {\it Phys.
Rev.} {\bf D53} (1996) 4775.}\nref\sg{T.~Gehrmann and W.~J.~Stirling
{\it Phys. Rev.} {\bf D53} (1996) 6100.}\refs{\bfra,\gr,\sg}
A detailed study of  the shape of polarized parton distributions\cite\sg\
shows (see Fig.~3) that while it is possible to separate out valence
(defined as $\Delta \bar q-\Delta q$) and sea components (assumed to
be SU(3) flavor symmetric) the constraints on the gluon distribution
are only relatively loose at large $x$
($x\gsim 0.1$) where scaling violations
driven by it are very small: in particular, the data disfavor but cannot rule
out a change of sign of $\Delta g$ at large $x$.
There is, however, remarkable agreement
between the shape of $\Delta g$ obtained by different groups under
rather different assumptions.\ref\giorom{G.~Ridolfi, {\tt hep-ph/9610214}.}

\topinsert
\epsfysize=6truecm
\vskip4truecm
\hskip-7truecm\hfil\epsfbox{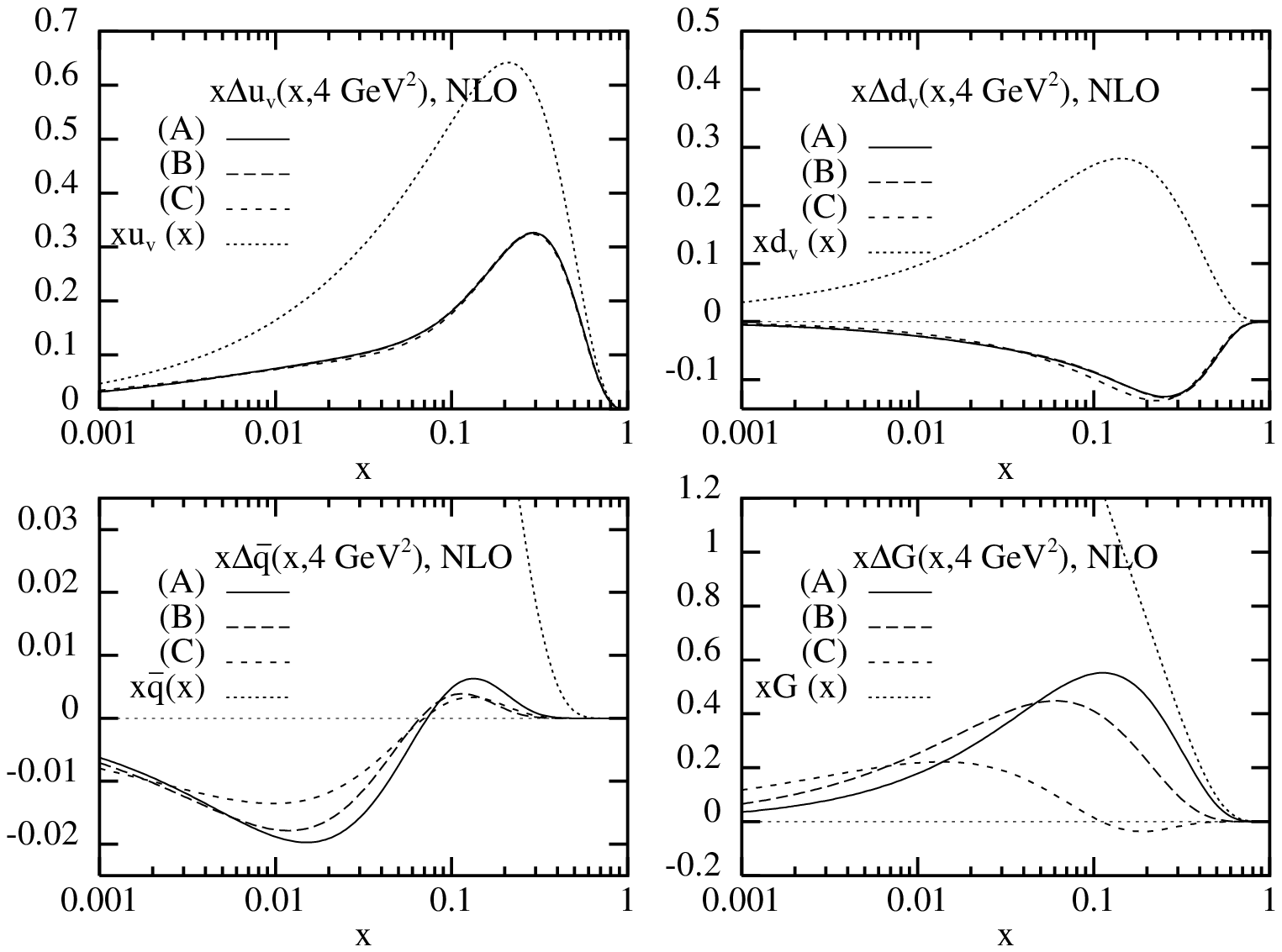}\hfil
\vskip-3.5truecm
\bigskip\noindent
{\abstractfont\baselineskip=10truept
\hskip-8pt Fig. 3: Polarized parton distributions (from
Ref.~\sg). The sets A-C correspond to different choices for the large $x$
behavior of $\Delta g$; the dotted curves are unpolarized
distributions.\smallskip}
\vskip-.5truecm
\medskip
\endinsert
The small $x$ behavior of the parton distributions
is particularly interesting. The proton data (Fig.~4a) display
a marked rise at small $x$ (and low $Q^2$, where the small $x$ data are taken):
 since $g_1^p$ at small $x$
is positive,
this appears to be due to a rise of the nonsinglet contribution,
because a rising singlet would produce substantial scaling violations
which turn it into a rapid drop. Indeed, the data for the deuteron, which
is almost pure singlet (the octet contribution to it
is quite small) do not display such a rise (Fig.~4b). This then
implies\refs{\bfra,\sg} an
effective behavior $\Delta q_{\rm NS}\sim {1\over\sqrt{x}}$ in the
measured region, in contradiction to Regge expectations, and reminescent
of the predictions\refs{\kl,\ermns}
based on the summation of double logs discussed
in Sect.~2.\footnote*{\footnotefont\baselineskip 10 truept
Interestingly, these predictions also imply that the small $x$ behavior
of the polarized and unpolarized nonsinglet should be similar; this also
appears to agree with the data.
\vskip1pt} Much more precise data at smaller values of $x$ will be required
to elucidate this issue.

The first precision data on the neutron structure function $g_1^n$
(from scattering on a $^3$He target, see sect.~5 below)
have recently been released in preliminary
form.\ref\slache{Y.~G.~Kolomensky, for the E154 Collab.,
 talk at SPIN '96, Amsterdam,
September 1996.} Comparison with these data of the predicted $g_1^n$,
based on parton distributions derived\cite\newfit\ from fits
to previous data  shows
excellent agreement (Fig.~4c), demonstrating that
our current understanding of the triplet/singlet separation
and of the small $x$ behavior of the nonsinglet are reasonably good.

\smallskip
\subsec{First moments}
\topinsert
\vskip-3.5truecm\vbox{
\hbox{\hskip-1.7truecm
\hfil\epsfxsize=9truecm\epsfbox{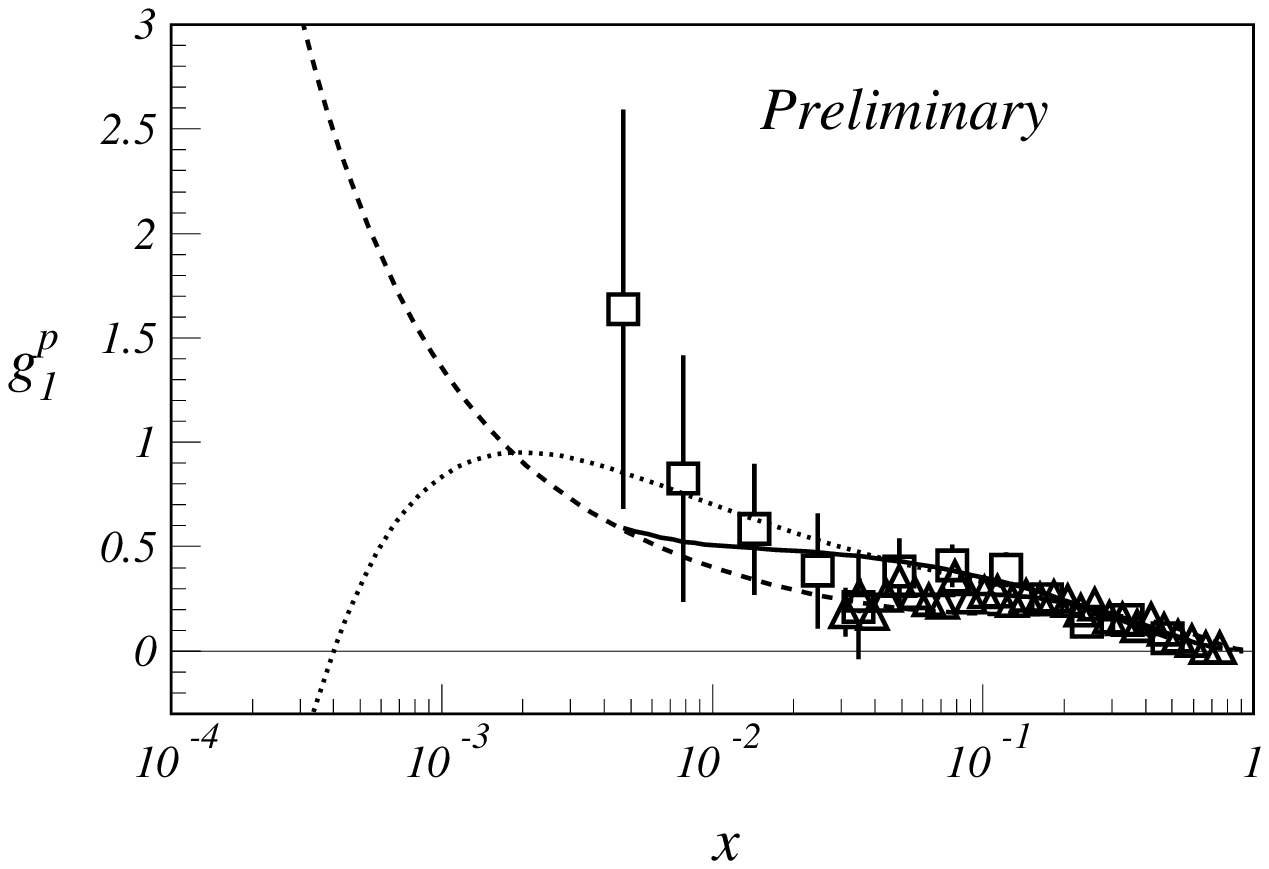}\hskip-2.truecm
\epsfxsize=9truecm\epsfbox{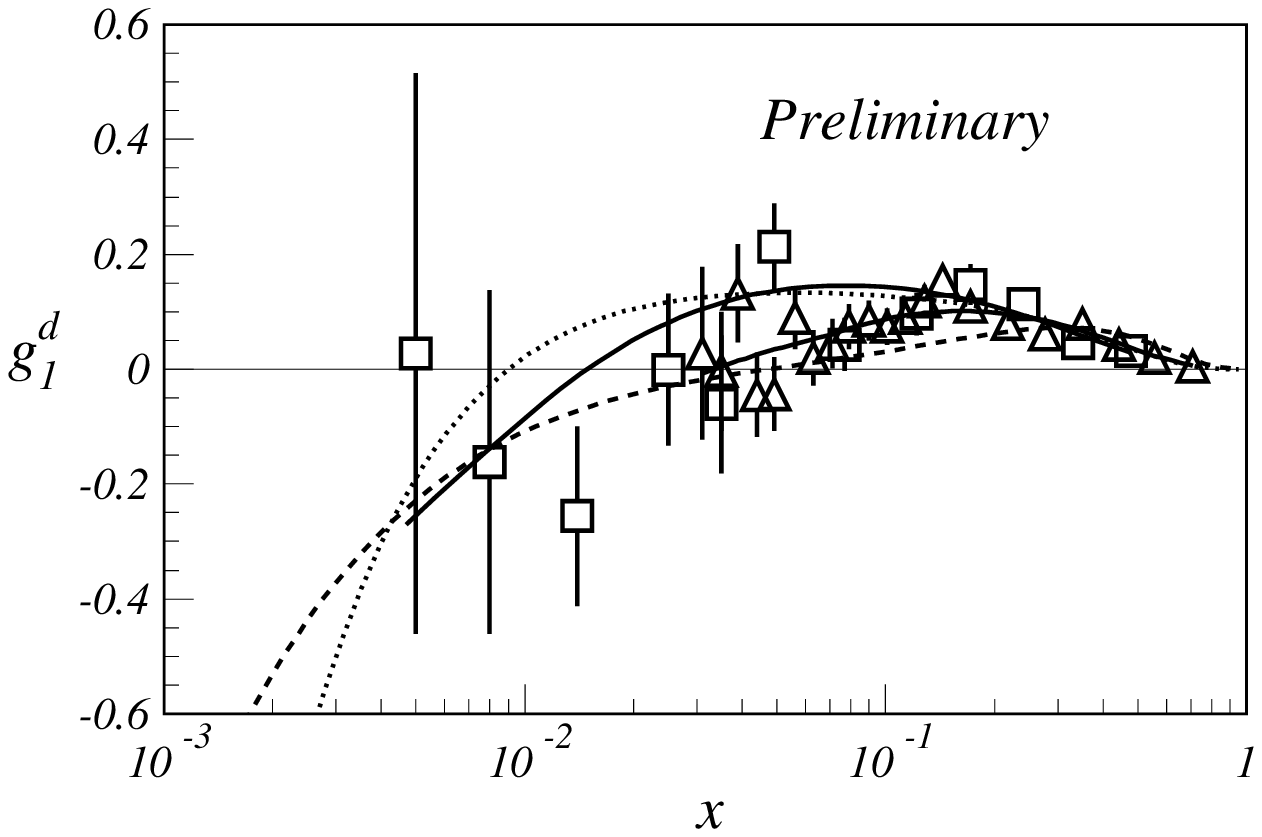}\hfil}}
\vskip-4.truecm
\epsfxsize=8truecm
\vskip-3.truecm
\hfil\epsfbox{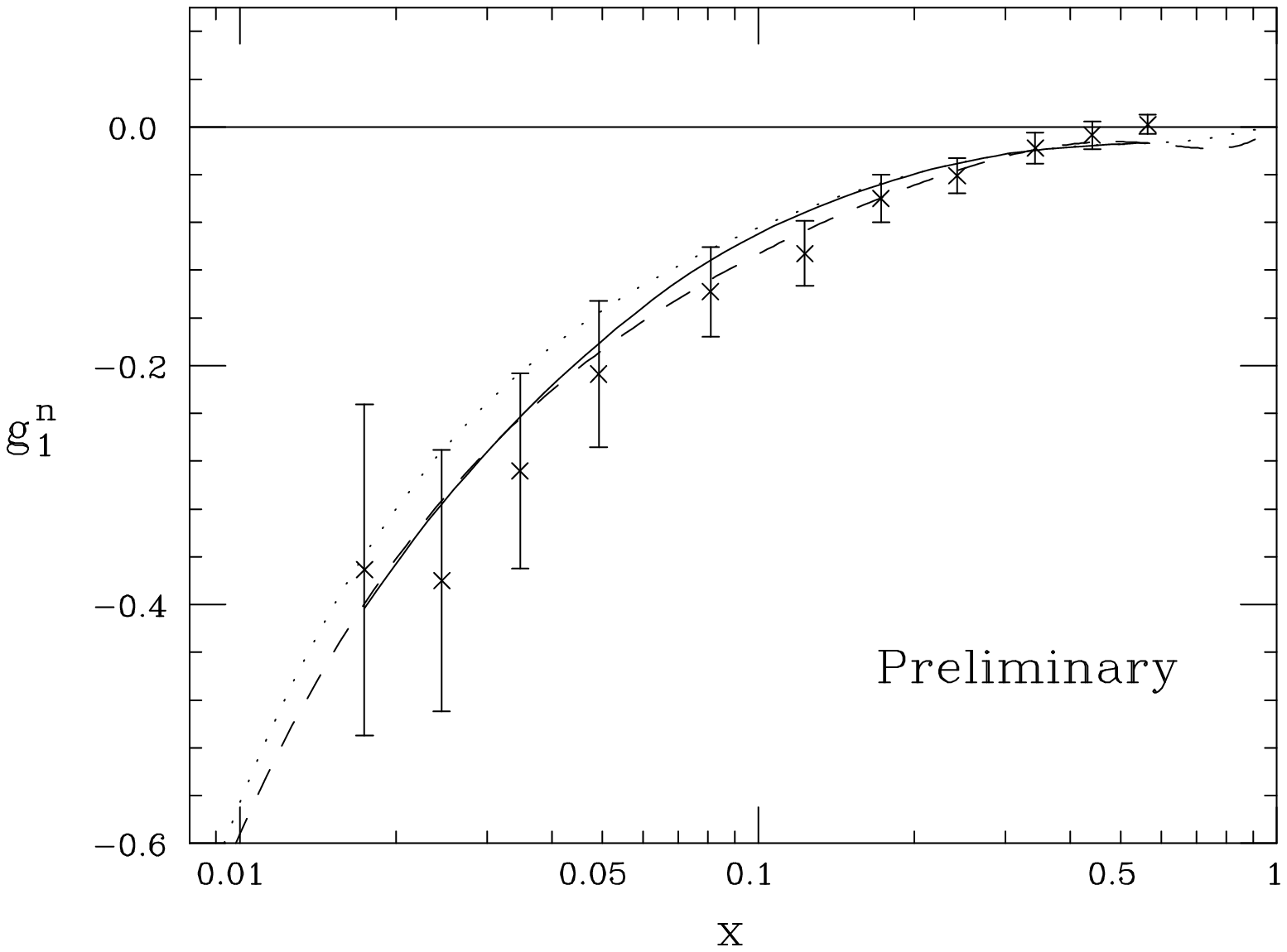}\hfil
\vskip-3.truecm
\bigskip\noindent{\abstractfont\baselineskip=10truept
\hskip-8pt
Fig. 4: Plot of $g_1$ for (a) proton, (b) deuteron and (c)
neutron.
The proton data are as in
Fig.~2,
the deuteron data are from Ref.~\xref\smcrm\ (squares, preliminary) and
from Ref.~\xref\slacd\ (triangles) and the neutron data are from
Ref.~\xref\slache. The curves  are a NLO
fit\cite\newfit\ (as in Fig.~2) to the data of (a) and (b) only.}
\vskip1pt
\vskip-.3truecm
\smallskip
\medskip
\endinsert
The NLO analysis can in particular be used\cite\bfra\
to give a precision determination of the
first moment of $g_1$, which is of primary theoretical interest because
of its relation to the singlet axial current:
\eqn\sing
{\Gamma^{\rm S}_1(Q^2)\equiv\int_0^1\!dx\,g^{\rm S}_1(x,Q^2)={\langle
e^2\rangle\over 2}
C_{\rm S} [\alpha_s(Q^2)] a_0(Q^2),}
where $a_0$ is the singlet axial charge
$a_0(Q^2)=\langle p,s|{j^\mu_5}|p,s\rangle
{s_\mu\over M}$. Unlike its nonsinglet counterpart eq.~\tripchar,
$a_0(Q^2)$ depends on scale (see Fig.~5) because the classical conservation
of the singlet axial current is spoiled by the axial anomaly in the
quantized theory. Its nucleon matrix element measures a nontrivial
combination [eq.~\gone] of the first moments of the quark and gluon
distributions (the total quark and gluon spin),
which at LO simply reduces to
$a_0(Q^2)=\Delta \Sigma(1,Q^2)$.
 The first moment is usually extracted from
the data by assuming $Q^2$ independence of the asymmetry $A_1=g_1/F_1$
to evolve all data to a common scale,\ref\elkar{J.~Ellis and M.~Karliner,
{\it Phys. Lett.}
{\bf B313} (1993) 131.} and extrapolating the data to $x=0$
on the basis of the Regge expectation.\cite\heim\
The axial charge can then be obtained
directly by using SU(3) symmetry to separate off the nonsinglet
component, and the perturbative expression of $C_{\rm S}(\alpha_s)$
(which is known\ref\lars{S.~A.~Larin, {\it Phys}. {\it Lett}.
{\bf B334} (1994) 192.} up to order $\alpha_s^2$). The
results thus found
from the published (or preliminary) values of the first moment,
evolved to a common scale of 5~GeV$^2$,
are collected  in the table.\footnote*{\footnotefont\baselineskip 10 truept
We used the $O(\alpha_s^3)$ expression\cite\larns\ of the nonsinglet
and the $O(\alpha_s^2)$ expression\cite\lars\ of the
singlet coefficient function with\cite\guias\ $\alpha_s(M_z)=0.119\pm 0.006$,
and the values of the triplet and octet
charges from Ref.~\nref\cloro{F.~E.~Close and R.~G.~Roberts,
{\it Phys. Lett.} {\bf B336} (1994) 257.}\xref\cloro.\vskip1pt}

Neither of these evolution and extrapolation procedures is
consistent with perturbative QCD, as discussed in Sect.~2: the information
contained in the data can instead be consistently extracted from a global
NLO fit. In such case, the primary quantities which enter the analysis are
the first moment of the quark and gluon distribution, from which the
first moment of $g_1$ is then extracted using eq.~\gone, and the
singlet charge using eq.~\sing. The main result of such an
analysis\cite\bfra\ is that
the naive procedure systematically  underestimates the
value of $a_0$, and substantially underestimates the theoretical
uncertainty on it. This is partly due to the fact that the small $x$
behavior is overconstrained by the Regge ansatz, and partly to the
fact that corrections due to perturbative evolution
(which are only known up to NLO) are potentially large.
In the NLO analysis the small $x$ behavior is fitted to the data and
constrained by assuming a reasonably smooth form of the starting parton
distributions, while higher order corrections to evolution are estimated
by varying the renormalization and factorization scale. Using
the data of Ref.s~\xref\smcp--\xref\slacd\   leads to
the value\cite\bfra
\eqn\salamino{a_0(5~{\rm GeV}^2)=0.15
\epm{0.16}{0.11}}
of the axial charge, to be compared to those of the table.

\nref\herm{D.~De~Schepper, for the HERMES Collab.,
talk at
``DIS 96'', Rome, April 1996.}
\topinsert\hfil\hskip-1.truecm
\vbox{\tabskip=0pt \offinterlineskip
      \def\tablerule{\noalign{\hrule}}
      \halign to 350pt{\strut#&\vrule#\tabskip=1em plus2em
                   &\hfil#\hfil&\vrule#
                   &#\hfil&\vrule#
                   &#\hfil&\vrule\hskip1pt\vrule#
                   &#\hfil&\vrule#
                   &#\hfil&\vrule#
                   &\hfil#&\vrule#\tabskip=0pt\cr\tablerule
      &&\omit\hidewidth Ref.\hidewidth
      &&\omit\hidewidth Targ.\hidewidth
      &&\omit\hidewidth $a_0(5~{\rm GeV}^2)$ \hidewidth
      &&\omit\hidewidth Ref.\hidewidth
      &&\omit\hidewidth Targ.\hidewidth
             &&\omit\hidewidth $a_0(5~{\rm GeV}^2)$\hidewidth&\cr\tablerule
&& \xref\smcp&&p && $0.27\pm 0.15$ && \xref\slacrm && p && $0.24\pm 0.09$
      &\cr\tablerule
&& \xref\slacp&&p && $0.31\pm 0.11$ && \xref\smcrm && d && $0.22\pm 0.09$
      &\cr\tablerule
&& \xref\smcd&&d && $0.16\pm 0.11$ && \xref\slacrm && $^3$He  && $0.37\pm 0.12$
      &\cr\tablerule
&& \xref\slacd&&d && $0.28\pm 0.06$ && \xref\herm && $^3$He  && $0.41\pm 0.22$
              &\cr\tablerule}}
\vskip-.5truecm
\hfil\bigskip\noindent{\abstractfont\baselineskip10pt\noindent\hskip-8pt
The singlet axial charge $a_0$ extracted from the
values of the first moment $\Gamma_1$ published by the experimental
collaborations, and evolved to a common scale. The data in
the last column are all preliminary.
 \smallskip}
\vskip-.2truecm
\endinsert
Experimental results for the singlet axial charge can be compared
to expectations based
on the quark model. In particular, the Zweig rule suggests that the
strange quark component of the nucleon should be small: this then
leads to the prediction that the singlet and the octet axial charge
should be approximately equal to each other (Ellis-Jaffe sum
rule\ref\elja{J.~Ellis and R.~L.~Jaffe, {\it Phys.
Rev.}
{\bf D9} (1974) 1444.}). Because
the octet axial charge is scale independent to all orders, while the singlet
charge, starting at NLO, is not,  this prediction can only be
exactly satisfied
at one particular scale. However, the scale dependence of $a_0$
is rather slight
even at relatively small  scales (Fig.~5b), so that the
Ellis-Jaffe prediction can be considered to be an approximate one,
which
the result eq.~\salamino\  would then strongly contradict (the
octet charge is\cite\cloro\
$a_8=0.579\pm0.025$). However, if one were to boldly extrapolate
the perturbative scale dependence of $a_0$ outside the
perturbative region proper (Fig.~5a), then
the sum rule could be satisfied by  assuming it to be valid
at a ``hadronic'' low scale of order of several hundreds MeV.

\topinsert
\vskip-2.5truecm\vbox{
\hbox{\hskip-.5truecm
\hfil\epsfxsize=7truecm\epsfbox{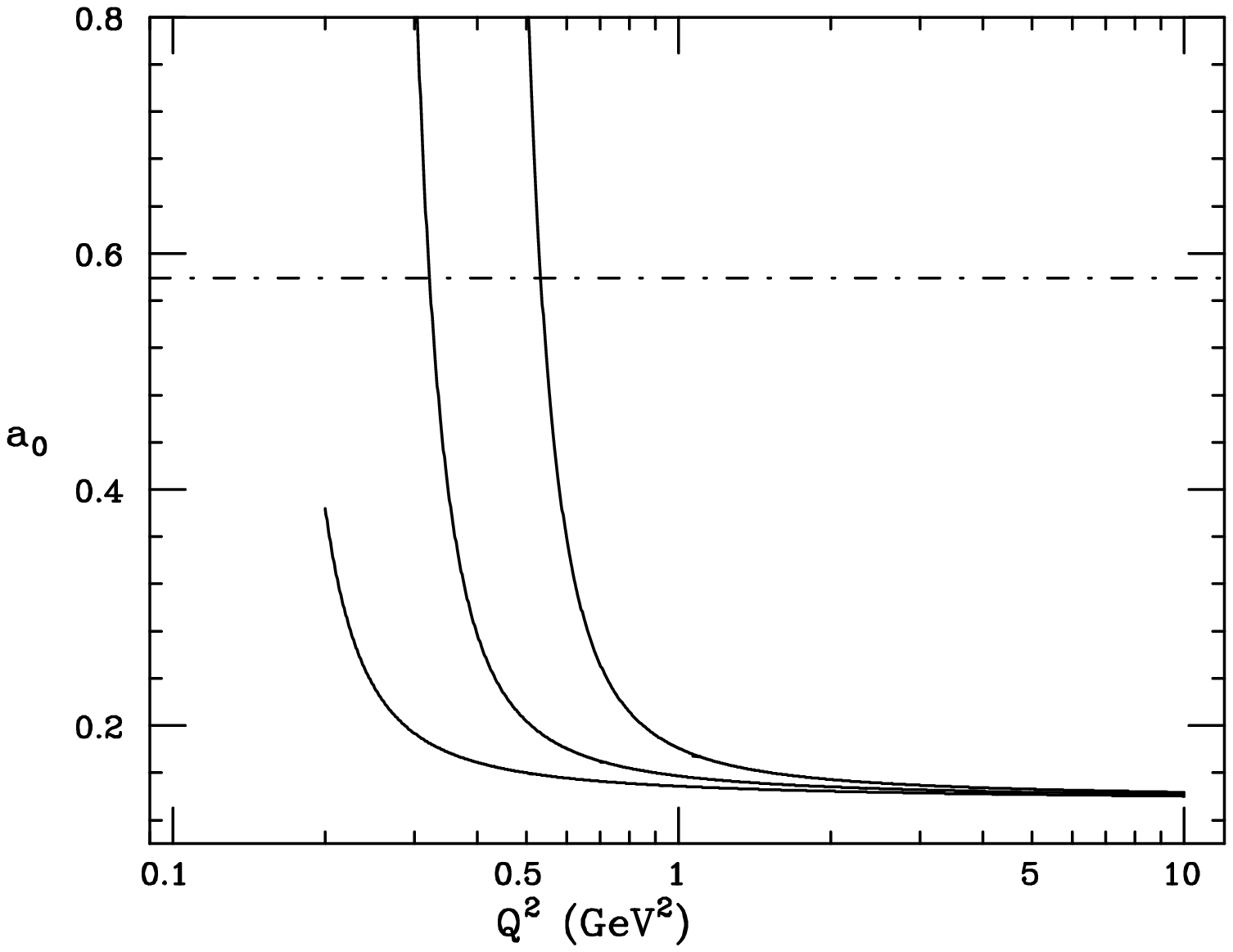}\hskip-2.truecm
\hskip.5truecm
\epsfxsize=7truecm\epsfbox{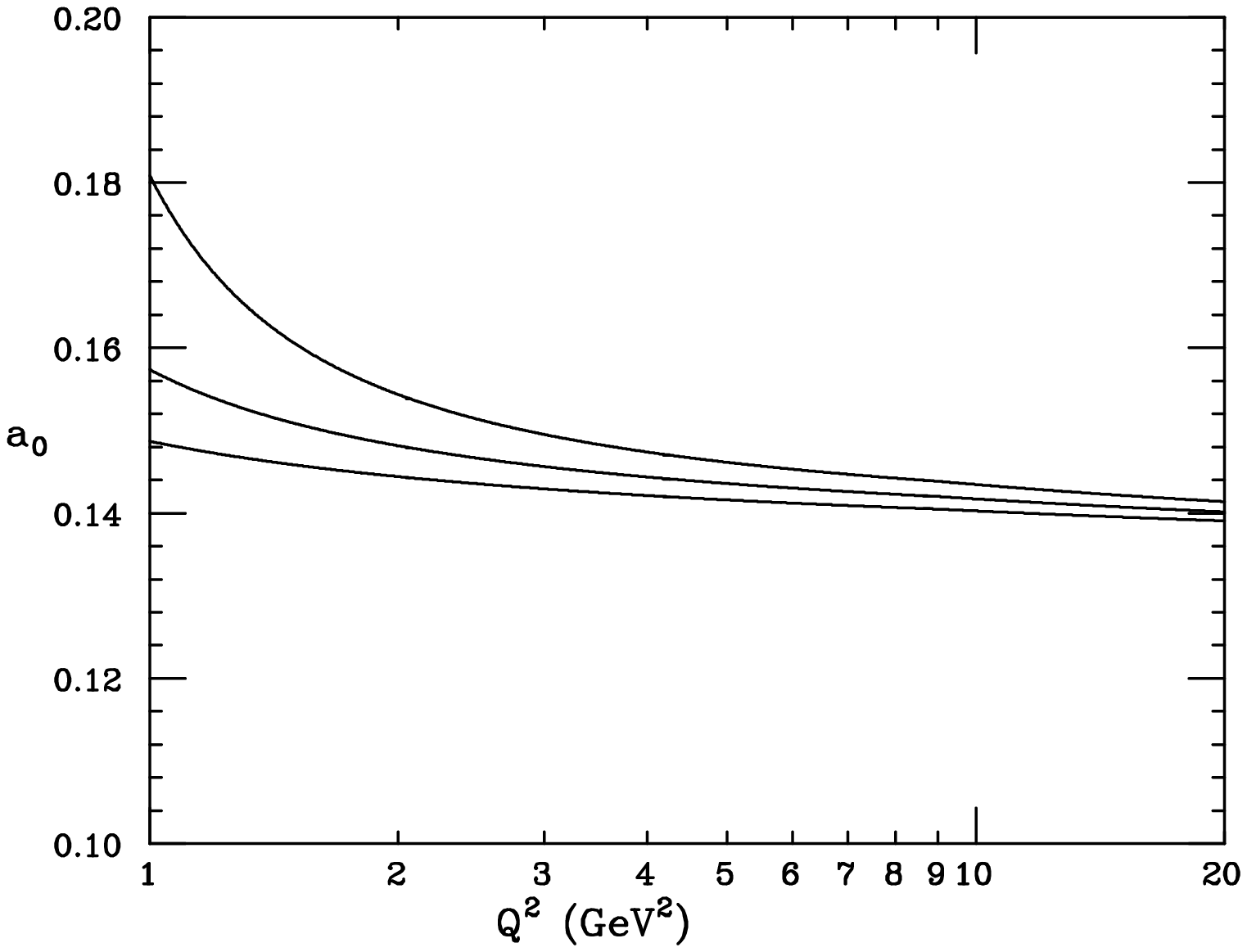}\hfil}}
\vskip-3.truecm
\bigskip\noindent{\abstractfont\baselineskip=10truept
\hskip-8pt
Fig. 5: Scale dependence of the axial charge $a_0(Q^2)$ eq.~\salamino\
computed at order $\alpha^2$ at low (a) and intermediate (b)
scales. The three curves
correspond to $\alpha_s(M_Z)=0.119\pm0.006$. The dot-dashed line is the value
of the octet charge $a_8$.
\vskip1pt}
\vskip-.3truecm
\smallskip
\medskip
\endinsert
Now, quark model predictions should presumably apply to partonic observables,
i.e. in this case to the
first
moments of the polarized  quark and gluon distributions.
However, the definition of the first
moment of the quark distribution is entirely
scheme dependent. This follows from the fact that
$\alpha_s\Delta g(1,Q^2)$ is  scale
independent at LO:\ref\alros{G.~Altarelli and G.~G.~Ross,
\PL\vyp{B212}{1988}{391}.}
if one performs a scheme change
$\Delta \Sigma^\prime(1,Q^2)=[1+\alpha_s z_{qg}\Delta g(1,Q^2)]
\Delta \Sigma(1,Q^2)$ then
$\Delta \Sigma^\prime(1,Q^2)-\Delta \Sigma(1,Q^2)$
is asymptotically constant, so
even asymptotically the ambiguity on $\Delta \Sigma$ is of the same order
as its size.\footnote*{\footnotefont\baselineskip 10 truept
Notice that this is not the case for $\Delta g(1,Q^2)$ itself: if
$\Delta g^\prime(1,Q^2)=[1+\alpha_s z_{gg}]\Delta g(1,Q^2)$
then even though $\Delta g^\prime(1,Q^2)-\Delta g(1,Q^2)$ is asymptotically
constant $\Delta g$ itself diverges (as $\alpha_s^{-1}$) in the same
limit, so the relative ambiguity vanishes asymptotically.
\vskip1pt} It is then possible in particular to choose a scheme where
$\Delta\Sigma(1)$ is scale independent. In this scheme one can meaningfully
compare the singlet quark distributions with the Ellis-Jaffe prediction since
then the first moments of all polarized quark distributions (and linear
combinations thereof) are separately scale independent.\cite\alros

As the NLO analysis actually derives the axial charge from the
polarized quark
and gluon distributions, and specifically their first moments,
these are also determined in the process.
Using a scheme\nref\absch{R.~D.~Ball, {\tt hep-ph/9511330};
S.~Forte, R.~D.~Ball and G.~Ridolfi, {\tt hep-ph/9608399}.}\refs{\bfra,\absch}
where $\Delta\Sigma(1)$
is scale independent leads to\cite\bfra\
\eqn\salamone{\Delta \Sigma(1)
=0.5\pm0.1,\qquad\Delta g(1,5\> \hbox{GeV}^2 )
=2.6\pm 1.4.}
An analysis\cite\sg\ with a different scheme choice for the quark leads
essentially to the same result for $\Delta g(1,Q^2)$.
This value of $
\Delta\Sigma(1)$ agrees
with the Ellis-Jaffe prediction;
the size of $\Delta g(1,Q^2)$ may
appear to be large, but is in fact natural recalling that it scales
as $1/\alpha_s$: if one were to use NLO evolution down to very low scales
the value eq.~\salamone\ would corresponds to vanishing
of $\Delta g(1)$ around $700$~MeV.
\goodbreak
\medskip
\newsec{\bf What can we learn?}
\nobreak
The present generation of polarized experiments has brought our
understanding of polarized parton distributions from the parton model
to the level where it makes sense to use QCD at NLO.
The data have shown evidence
for scaling violations (Fig.~2), have given us a first understanding
of the shape of parton distributions (Fig.~3,~4), and have led
to a determination of the singlet axial charge and the first moment
of the gluon distribution [eq.s~\salamino,\salamone] which, given
the uncertainties involved, should be considered surprisingly
precise.

While the determination of the first moment of the gluon distribution
is dominated by the statistics and will thus improve somewhat as more abundant
data
in the currently explored kinematic region become available
(specifically from SMC\cite\smcrm\ and E155\cite\slacrm),
the determination of the singlet axial charge is already dominated
by the systematics related to higher order corrections, as well as by the
uncertainty in the
small $x$ behavior. Only the availability of data with a more complete
coverage of the $(x,Q^2)$ plane (such as those from the
proposed GSI collider\ref\gsi{D.~von~Harrach, V.~Metag and A.~Sch\"afer,
{\tt http://www.th.physik.uni-frankfurt.de/$\sim$schaefer/gsi}.}
 or from HERA with a polarized proton beam\ref\polher{J.~Lichtenstadt, talk at
``DIS 96'', Rome, April 1996.})
would allow a substantial improvement
here, as well as a
 determination of higher moments and in general
of the $x$ dependence of parton distributions.
The study of scaling violations at small $x$
in particular should turn out to be very fruitful:  it could
confirm the violation of the Regge-based expectation
for the small $x$ shape of $g_1$ suggested by current data,
provide evidence for the non-Regge behavior eq.~\sxasy\ predicted
by perturbative QCD  (and
seen\ref\hera{S.~Forte and R.~D.~Ball, {\it Acta Phys. Pol.} {\bf B26} (1995)
2097; R.~Ball and A.~De~Roeck, {\tt hep-ph/9609309}. }
in unpolarized singlet structure functions at HERA), show the dramatic
change of sign of $g_1^p$ (see Fig.~4a)
predicted by the QCD evolution equations
as the scale is raised  (thereby leading to a  very precise determination
of the polarized gluon distribution\ref\heraws{R.~D.~Ball et al.,
{\tt hep-ph/9609515}}),
and possibly show evidence of the
double logarithmic effects which characterize the polarized small $x$
behavior.\refs{\ermns,\erms}
On the theoretical side, there is room for improvement from
 a better understanding of double logs at small
$x$, consistent with the running of $\alpha_s$, and from a
more detailed understanding of higher twist effects, both
kinematical\ref\kawa{H.~Kawamura and
T.~Uematsu, {\it Phys}. {\it Lett}. {\bf B343} (1995) 346.}
(such as related to target mass effects) and dynamical,\cite\lech\ which have
been so far studied systematically only at the level of first moments.

The theory and phenomenology which we have discussed so far are
based on fully inclusive measurements of longitudinal polarization
with lepton beams and nucleon (proton and deuteron, actually)
targets. These allow (at leading twist) only the determination of the
combinations of polarized parton distributions which contribute
to $g_1$ eq.~\gone. Firstly, this only measures
 the $C$-even combination $q+\bar q$ of polarized quark
distributions: the ``valence'' $q-\bar q$ component can only
be disentangled\refs{\gr,\sg} by making specific assumptions on the
symmetry of the sea. Furthermore, whereas comparing proton
and deuteron data allows the separation of the isotriplet component,
a further flavor decomposition can only be done at the level of first
moments by using SU(3) symmetry and data from hyperon $\beta$ decays,
while for the full $x$ dependence it again requires\refs{\gr,\sg,\bfra}
assumptions on the symmetry of the sea.

For all moments beyond the
first the singlet quark and gluon contributions can then separated by the
observation
of scaling violations.
The case of the first moment is special because,
whereas the gluon is again determined by scaling violations, the quark is
completely ambiguous (see Sect.~3), and can thus
only be determined within a specific scheme choice.
It is important to notice that this is an inevitable consequence of
the scaling properties of $\Delta g(1,Q^2)$, which allow
an asymptotically nonvanishing redefinition of the singlet
quark,  and it does not depend on the specific
process. Thus,   a determination of $\Delta\Sigma(1,Q^2)$
from any hard process will be subject to the same ambiguity, and
there is no piece of data which could eliminate it, even in principle.
Notice however that the singlet valence
component, which decouples from the gluon, is not subject to this
ambiguity.

A simple way of widening the source of available information is to consider
nuclear targets. In particular, $^3$He targets
provide a direct measurement of the neutron structure function,
to the extent that only the neutron carries the nuclear spin. Corrections
due to higher spin components in the nuclear wave
function\ref\ciofi{C.~Ciofi~degli~Atti et al, {\it Phys. Rev.}
{\bf C48 } (1993) 968}
have been studied and seem under theoretical control
(Fermi motion effects
appear\ref\mel{W.~Melnitchouk, G.~Piller and  A.~W.~Thomas, {\it  Phys. Lett.}
{\bf B346} (1995) 165.} to be negligible). On the contrary,
there are only preliminary studies\ref\fran{L.~Frankfurt,
V.~Guzey and M.~Strikman, {\tt hep-ph/9602301}.}
(in the nonsinglet sector only)
of shadowing effects, which suggest that they may be quite large
(up to 20\% around $x=0.1$ where $g_1$ is sizable): a systematic investigation
of these effects is required if data from the  $^3$He
experiments (in particular the statistically very precise
E154\cite\slache\ data of Fig.~4c)
 are to be seriously taken as a  precise determination of the neutron
polarized structure function.

More ambitious programs (perhaps more theoretically than experimentally so)
involve going beyond fully inclusive measurements. Tagging hadrons
in the final state of a DIS event leads to information on the struck
current, provided one can separate out the fragmentation of
the current itself (as described by fragmentation functions)
from the fragmentation of the target
(parametrized by fracture
functions\ref\luca{L.~Trentadue and G.~Veneziano, {\it Phys. Lett.}
{\bf B323} (1994) 201.}). In principle, measurements in the current
fragmentation region makes then possible
a separation into
individual flavor components, and also into valence and
sea contributions:\footnote*{\footnotefont\baselineskip 10truept
An alternative method to disentangle valence
and sea components for individual flavors is to use charged-current
events.\ref\paolo{M.~Anselmino, P.~Gambino and J.~Kalinowski,
{\it Z.~Phys.} {\bf C64} (1994) 267.} Even though this would be feasible
at HERA with a polarized proton beam, it does not appear to be competitive.
\vskip1pt} however, beyond LO there is no simple way of
associating simple observables (such as asymmetry ratios) to
individual matrix elements (such as the isotriplet valence component),
other than by making assumptions, for instance about the symmetry
of the sea, which is what one would want to test in the first place.
However, these data can be used in a global NLO analysis (the general
framework of which can be already set up\ref\arg{D.~de~Florian et al.,
{\tt hep-ph/9603302}.}) where they may significantly constrain the final
result. Such experiments are currently underway at
HERMES,\ref\hrmsi{K.~Ackerstaff, for the HERMES Collab., talk at
``DIS 96'', Rome, April 1996.} while an ambitious
program has been proposed at CERN.\ref\compass{COMPASS proposal, preprint
CERN/SPSLC 96-14.}
On the other hand, measurements of target fragmentation may provide
nonperturbative information, and specifically
allow\ref\grah{G.~M.~Shore, {\tt hep-ph/9609438}} to test whether
the smallness of the singlet axial charge discussed in sect.~4.2
is a peculiar property of the nucleon, or rather a target-independent
property of the QCD
vacuum.\ref\vesho{S.~Narison, G.~M.~Shore and G.~Veneziano,
{\it Nucl. Phys.} {\bf B433} (1995) 209.}

A separate class of exclusive or
semi-inclusive experiments can be used to measure
the polarized gluon distributions in processes
where it contributes at leading order (such as heavy quark
production by photon-gluon fusion). It is then possible
to evade the  tradeoff of inclusive DIS, where the
gluon is measured by scaling violations, which are larger at low
scales where however scheme ambiguities are also large; even though there
are then other theoretical difficulties characteristic of
non-inclusive processes (such as scale ambiguities). The cleanest
measurement here is possibly heavy  quark
photoproduction, where
since only one large scale (the quark mass) is present
the cross section can be reliably computed in
perturbation,\ref\glurcc{M.~Gl\"uck, E.~Reya and W.~Vogelsang,
{\it Nucl. Phys.} {\bf B351} (1991) 579.}
so that a study of the $p_T$ dependence of the production
asymmetry can lead to a sensitive determination of the
gluon distribution.\ref\mastro{S.~Frixione and G.~Ridolfi,
{\tt hep-ph/9605209}.} A NLO computation of this process, which
is still lacking, would be highly desirable.

As more precise data become available, it will also be possible to take
the ambitious step of going beyond longitudinal
polarization. So far, the structure function $g_2$ has been considered
a higher twist background to the measurement of $g_1$,
and recent determinations\ref\gte{SMC
Collab., {\it Phys}. {Lett}. {\bf B336} (1994) 125;
E143 Collab., {\it Phys. Rev. Lett.} {\bf 76} (1996) 587.}
of it have essentially achieved the aim of reducing this background.
A study of the $x$ and $Q^2$ dependence of $g_2$ as might be
possible in a dedicated experiment would open Pandora's box of
twist-3 operators and their mixing,\ref\jafre{For a review
see R.~L.~Jaffe, {\tt hep-ph/9602236}.}
 and lead beyond the safe context
of parton distributions defined by twist two operators.


\nobreak
\smallskip
\noindent{\bf Acknowledgements:} I thank
V.~Hughes and the SMC
collaboration   and P.~Bosted (of the
E143 collaboration)  for
valuable information  on their data,
M.~Anselmino for discussions,
A.~Deshpande,  J.~Lichtenstadt and especially G.~Ridolfi
for help in the data analysis,
R.~Ball for a critical reading of the manuscript, and J.~Domingo and
K.~de~Jager for their hospitality in Williamsburg and Amsterdam.
\goodbreak

\immediate\closeout\rfile\writestoppt
\bigskip
\noindent{{\bf References}}\smallskip{\frenchspacing%
\parindent=20pt
\ninepoint\baselineskip=10pt
\escapechar=` \input refs.tmp\vfill\eject}\nonfrenchspacing
\vfill
\eject

\bye